\documentclass{article1}
\usepackage[latin2]{inputenc}
\usepackage{natbib1}
\usepackage[dvipsone]{epsfig}
\RequirePackage{epsf,graphics,leqno}

\def\newblock{\hskip .11em plus .33em minus .07em}

\LRH{G. Iv\'an and V. Grolmusz}

\RRH{Dimension reduction in bioinformatics}

\title{On Dimension reduction of clustering results in bioinformatics}

\authorA{G\'abor Iv\'an and\newline Vince Grolmusz*}

\affA{Eötvös Loránd University,\\ PIT Bioinformatics Group\\ and Uratim Ltd., 1117 Budapest, Hungary\\ E-mail: hugeaux@cs.elte.hu\\ E-mail: grolmusz@pitgroup.org
\newline $^{*}$Corresponding author }

\title{Dimension reduction of clustering results in bioinformatics} 
\author{Gábor Iván\,$^{\rm a, b}$, Vince Grolmusz\,$^{\rm a, b}$\footnote{to whom correspondence should be addressed}
\\
\small $^{\rm a}$ PIT Bioinformatics Group, Eötvös University,\\
 \small Pázmány Péter stny. 1/C, H-1117 Budapest, Hungary\\
\small $^{\rm b}$ Uratim Ltd.,  H-1118 Budapest, Hungary}

\begin{document}

\maketitle

\begin{abstract}
OPTICS is a density-based clustering algorithm that performs well
in a wide variety of applications. For a set of input objects,
the algorithm creates a so-called \emph{reachability plot} that
can be either used to produce cluster membership assignments,
or interpreted itself as an expressive two-dimensional representation
of the density-based clustering structure of the input set, even if the input set is embedded in higher dimensions.
The main focus of this work is a visualization method that can be used to assign colours
to all entries of the input database, based on hierarchically represented
\emph{a-priori} knowledge available for each of these objects.
Based on two different, bioinformatics-related applications we
illustrate how the proposed method can be efficiently used
to identify clusters with proven real-life relevance.
\end{abstract}


\maketitle

%%%%%%%%%%%%%%%%%%%%%%%%%%%%%%%
\section{Introduction}
%%%%%%%%%%%%%%%%%%%%%%%%%%%%%%%

Clustering algorithms are a useful branch of data mining techniques that
assign similar objects to the same group, creating several so-called ''clusters''. The method is a core technique in datamining and has numerous applications in bioinformatical datamining of 
biological sequence data \cite{Goncalves2012,Vertessy1994,Fiser2000,Ivan2007}, in biomedical image processing and analysis \cite{Rojas2011}, in proteomics data analysis \cite{Leung2012}, in microarray data analysis \cite{Kossenkov2010,Zhang2008}, protein-protein interaction network analysis \cite{Yang2007,Ivan2011} and phylogenomic analysis \cite{Leblond2010}.
Usually, the only requirement to find clusters is the existence of a similarity
measure that assigns a numeric value to the similarity of any object-pair.
Interpreting the output (and even properly setting the input parameters) of a given
clustering algorithm is task usually requiring much consideration. 

In this
work we propose a visualization method that makes use of hierarchically
represented \emph{a-priori} knowledge available about the input objects, and assigns
colours to them based on this information. We then show how the proposed method can help
identifying clusters with real-life relevance using the OPTICS clustering algorithm \cite{OPTICS, Ivan2009, Ivan2010}.

Any novel application of a bioinformatics method needs to be validated by detailed comparisons of known techniques and {\em a priori} knowledge (c.f.,\cite{pdbmend, Devill-Delun}). Our presented method yields a framework for this comparison: two independent clusterings can be superimposed: one clustering by the concave regions of OPTICS reachability diagram, and the other, completely independent clustering by the coloring of the data items in the reachability diagram. We applied this technique first in \cite{Ivan2009}, where we found that in protease enzyme families, the configuration of just four spatial points in these enormous protein structures definitively imply their exact enzymatic role. In the present work we formulate the visualization method itself.

The source code of the visualization algorithm with sample output is available at http://uratim.com/appendix\_visual\_article.zip.

\section{Overview of the OPTICS clustering algorithm}

Our visualization method proposed in the next Sections can be applied
to the output of any clustering algorithm. However, the usefulness of the method
is going to be presented using results of the specifically chosen OPTICS algorithm \cite{OPTICS},
as the simultaneous use of OPTICS and the hereby presented visualization technique
brings some further advantages.
First we give a brief description of the OPTICS clustering algorithm, and also the
justification of using this particular algorithm as a candidate to test our
proposed visualization method.

For data clustering we intended to use an algorithm that is capable of identifying
outlier points (also referred to as ''noise'') and is not biased towards even
sized or regular shaped clusters. Density-based clustering algorithms have these
desirable properties. The density of objects can be defined with a radius-like
$\epsilon$ parameter and an object-count lower limit ($minpts$): a neighbourhood
of some object $o$ is considered dense if there exist at least $minpts$
objects within a less-than-$\epsilon$ distance. As the clustering structure of
many real-data sets cannot be characterized by one (global) density parameter,
it seems advisable to eliminate one of the above two input parameters and
use it on the output instead.

The OPTICS (\emph{Ordering Points To Identify the
Clustering Structure}, \cite{OPTICS}) algorithm achieves this by
\emph{ordering} the objects contained in the database, creating
the so-called \emph{reachability plot}. The reachability plot is
generated by assigning a value called \emph{reachability distance}
to all the objects of the database, while processing the objects
in a specific order: the algorithm always chooses the object
reachable with the smallest possible $\epsilon$ distance while
maintaining the lower limit defined by $minpts$, meaning roughly
the ''most dense direction''. This ensures that the hierarchical
clustering structure of the database is also preserved.

The measure of local density for each object encountered is depicted
on the reachability plot that contains almost all the information
about the clustering structure of the database, although it does
not directly assign the objects to clusters. There exist several
methods that assign cluster memberships to objects based on the
OPTICS reachability plot; these may be of interest in a future
study. However -- with the proposed visualization method -- it is
possible to obtain quite usable results without even assigning any
particular cluster memberships to the objects: when using the OPTICS
clustering algorithm together with a specific similarity measure, we
would usually like to know whether the ''deep'' regions of the
reachability plot -- these are ''potential'' clusters --
correlate with some \emph{a priori}-known information.

\begin{figure}[!h]
\centering
\includegraphics[width=5.2in]{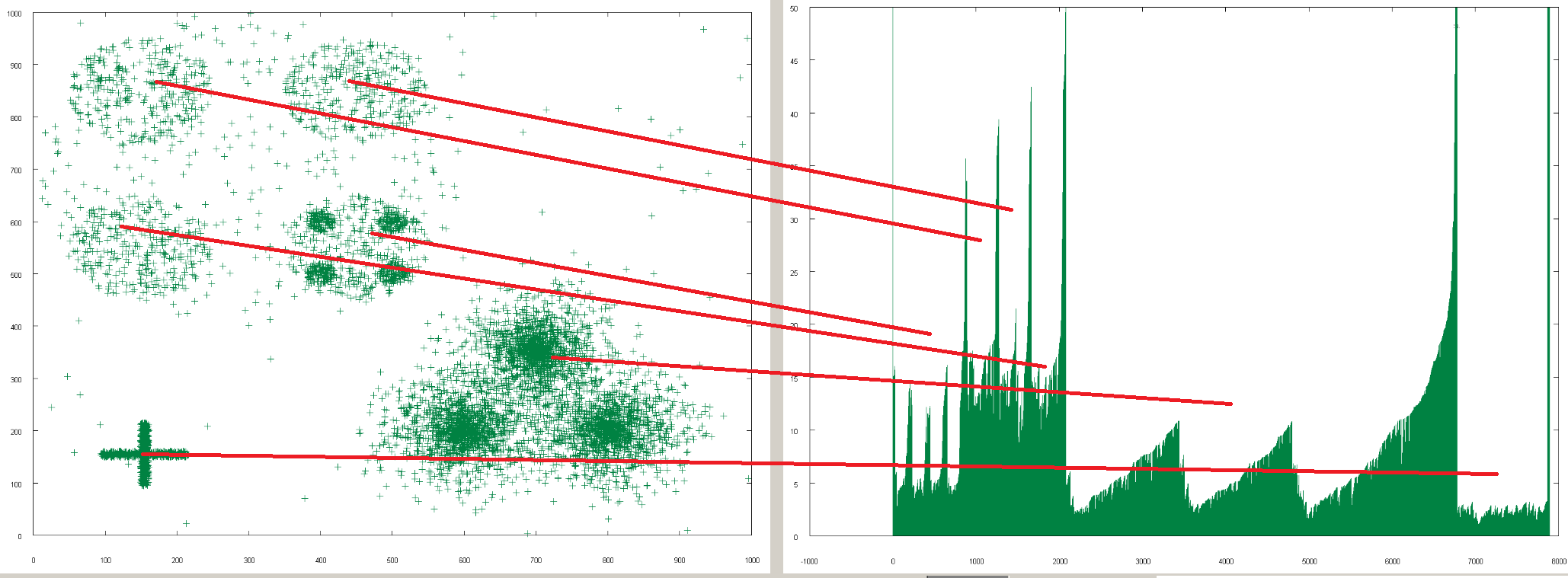}
\caption{OPTICS reachability plot (right) of a sample database (left) consisting
of clusters with different shape, size and density. The correspondence is denoted by red lines: note that clusters correspond to concave regions 
on the right, and hierarchical clusters are also clearly visible on the reachability plot: shallower regions containing deeper concavities correspond to hierarchical clusters. \label{figOPTICSexample1}}
\end{figure}

The reachability plot of some points scattered on a two-dimensional plane is
depicted on Figure \ref{figOPTICSexample1}. The applied similarity measure is
simply Euclidean distance. It is important to notice that the
OPTICS algorithm is capable of creating the reachability plot for objects
represented in arbitrary dimensions; it is only the similarity
measure that has to be changed accordingly.

As a side effect, OPTICS reduces the dimensionality of the input dataset; combining OPTICS
with the visualization method proposed later can be thus also used to visually
compare two hierarchical clusterings of (possibly) multi-dimensional datasets. The literature of dimension reduction
and visualization of high-dimensional data sets is quite rich
(e.g. \cite{VIS-EXAMPLE1}), which is also true for visualizing hierarchical
clusterings (e.g. \cite{VIS-EXAMPLE2}). Our method combines dimension reduction
with visualization, making it possible to compare clustering results
to an \emph{a-priori} given hierarchical classification without assigning
objects to specific clusters.

\section{Colouring nodes of the \emph{a priori}-known hierarchical data structure}

In the visualization phase we are going to
assign colours to each entry occurring on the x-axis of the OPTICS reachability plot,
based on the \emph{a-priori} given hierarchical classification of these objects.
The main idea is that we would like to use similar colours on entries
that belong to ''similar'' classes in the \emph{a priori}-known hierarchical
data structure. As this hierarchical structure can be conveniently represented
by a (non-binary) tree (a \emph{dendogram}), our aim is to assign colours to tree
nodes so that nodes having a short path between them (i.e. their common ancestor
is close to them) are assigned similar colours. We would also like to achieve
that the depth of a given node in this tree is somehow reflected in its colour.

We will use the HSB (Hue, Saturation, Brightness) representation
of colours. HSB coordinates can easily be converted to RGB (Red,
Green, Blue) colour coordinates. As an example, it is easy to see that
points of the whole Hue scale may be assigned to leaf nodes with full Saturation
and Brightness, while the root node may be coloured black. It is also straightforward
to use Saturation and Brightness as an indicator of depth (distance from root)
in the tree. This is the main motivation of the chosen HSB colour
coordinate system.

More precisely, assignment of colours to the
\emph{a-priori} given dendogram is carried out as follows.

\subsection{Determining Saturation and Brightness}

Saturation and Brightness assigned to a given tree node will be
(not necessarily directly) proportional to the level (distance from
root) of this particular node. The level of each node can be --
for instance -- determined by BFS (Breadth-First Search) algorithm
starting from the root. The root itself will always be black and
nodes on the lower levels will be ''more colourful'' -- see Figure \ref{fig1}.

\subsection{Determining Hue}

Let us call a tree node with no children a \emph{leaf}.
Let $C^t_k$ denote the $k$-th child of tree-node
$t$ (we may define an arbitrary ordering on child nodes).
Let the the weight of a leaf $l$ be $W(l) = 1$, and the
weight of a given node $t$ having $c$ children be the number
of nodes its subtree contains, or (given by a recursive definition):

\begin{equation} \label{eq:TreeColouring1}
W(t) = 1 + \sum_{k=1}^c W(C^k_t).
\end{equation}

Each node $t$ is assigned a closed hue interval $I(t) = [I^t_l, I^t_r]$;
let us denote the size of $I(t)$ with $S(t) = I^t_r - I^t_l$.
The root of the tree is assigned the whole hue interval $I(root) = [0; 1]$.
The hue assigned to any node $t$ is $H(t) = \frac{I^t_l - I^t_r}{2}$
(e.g. $H(root) = 0.5$).

The children of node $t$ are assigned non-overlapping subintervals
of $I(t)$ that are separated from each other. The latter means that
a certain part of $I(t)$ is not used at all at any child of $t$;
this unused part provides separator intervals to ensure
that Hues are highly distinct between nodes with different ancestors
(this is the motivation of using such separator intervals).
Let us denote the proportion of this unused hue interval to
$S(t)$ with $E$ (which is a global parameter of the colouring method).

A certain part of $I(t)$ is divided between the children of $t$.
The size of the hue interval assigned to the $k$-th child of $t$, $S(C^t_k)$ is

\begin{equation} \label{eq:TreeColouring2}
S(C^t_k) = \frac{W(C^t_k)}{W(t)-1} \cdot (1-E) \cdot S(t).
\end{equation}

In other words this means that we take the interval belonging to
node $t$ with size $S(t)$, and assign an $(1-E)$-proportion of this
interval to child nodes' hue intervals, and an the remaining will
be used to provide separator intervals that are not assigned to any
descendant of $t$. Sizes of assigned hue intervals are
proportional to the weight of the given child node.

We would now like to formulate the hue interval belonging to the
$C^t_k$. Let $c$ denote the number of children of node $t$. The
$c$ hue intervals determined for each child node $t$ are
separated by $c-1$ separator-intervals, each sized $\frac{E \cdot S(t)}{c-1}$.
Taking into account that the hue interval belonging to the $k$-th child
is preceded by $k-1$ equal-sized separators and $k-1$ hue intervals belonging to
the previous $k-1$ children,

\begin{equation}
I(C^t_k) = \left[ (k-1) \cdot \frac{E \cdot S(t)}{c-1} \cdot \sum\limits_{i=1}^{k-1}
S(C^t_k)\; ; \;\; (k-1) \cdot \frac{E \cdot S(t)}{c-1} \cdot \sum\limits_{i=1}^{k}
S(C^t_k) \right].
\end{equation}

\begin{figure}[!h]
\includegraphics[width=3.5in]{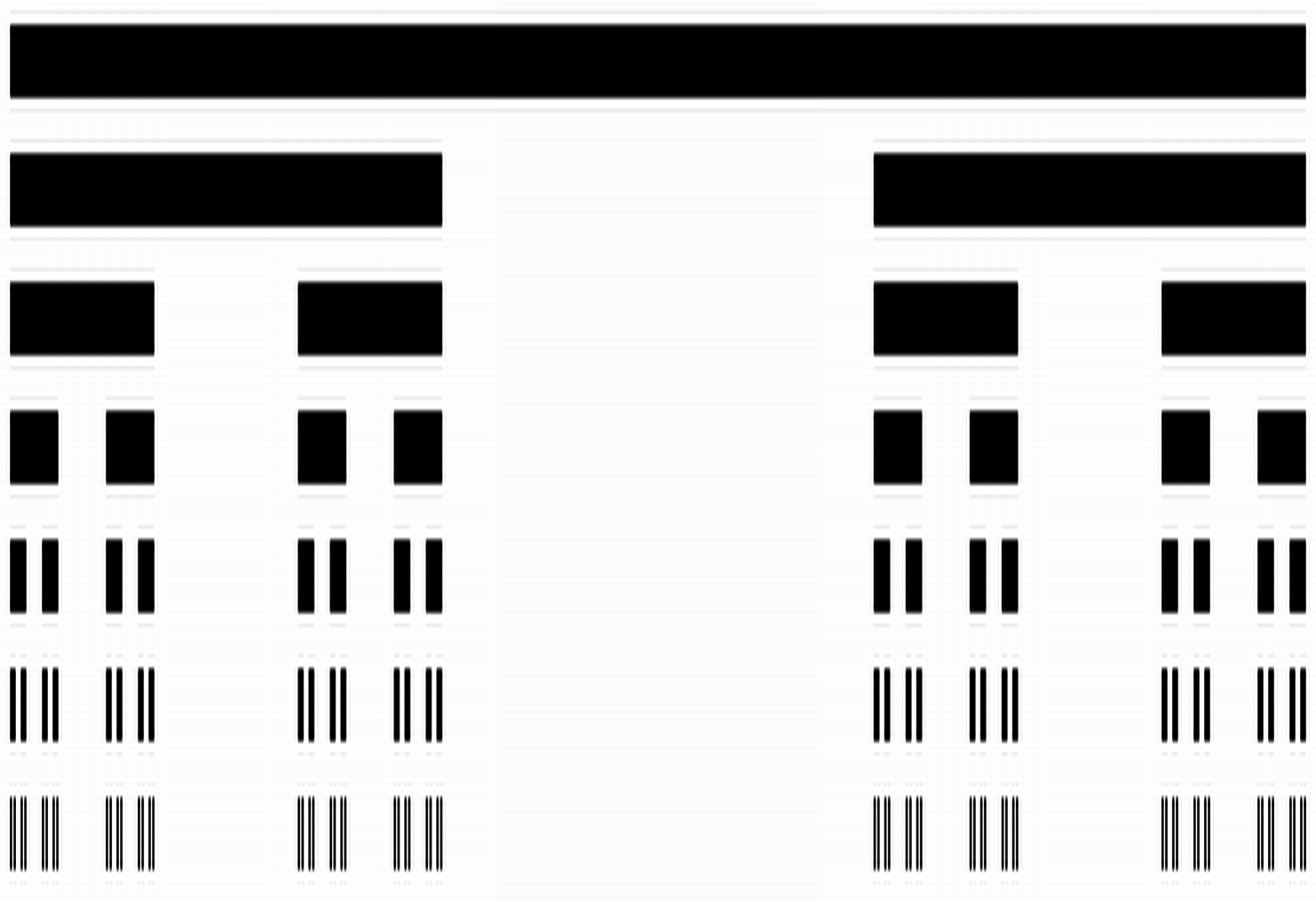}
\caption{Illustration of the sizes of Hue intervals occurring on the
levels of a binary tree ($E = \frac{1}{3}$, the $i$-th level has $2^i$ nodes) 
(Source: http://en.wikipedia.org/wiki/Cantor\_set, downloaded January 7, 2012).\label{fig2}}
\end{figure}

To illustrate the above principles, we provide two simple
examples. If we set $E = \frac{1}{3}$ and use a balanced
binary tree to be coloured, the hue intervals occurring at
each level of the tree will converge to the Cantor set if
we move farther and farther from the root (see Figure \ref{fig2}).

\begin{figure}[!h]
\includegraphics[width=4in]{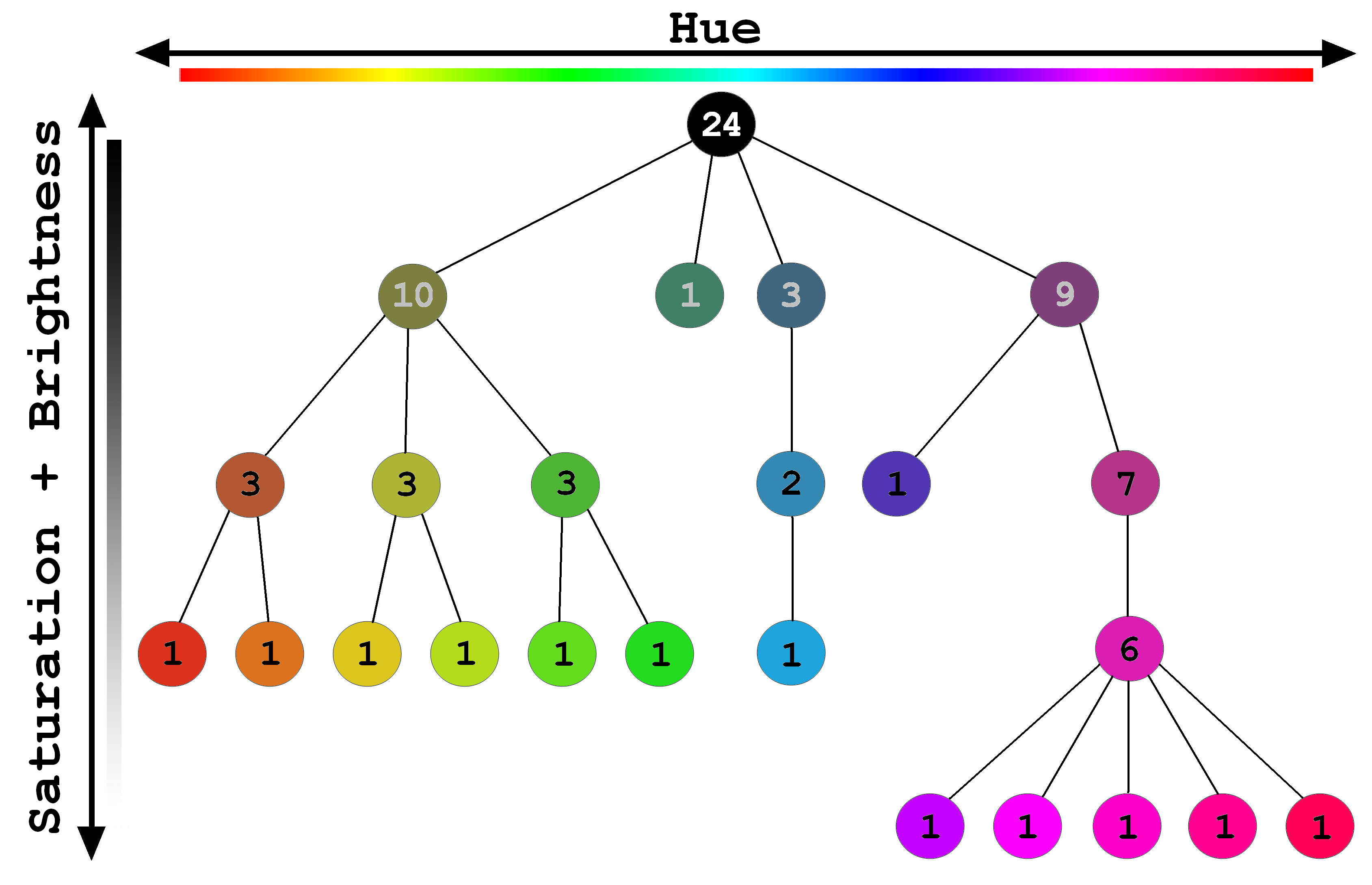}
\caption{Illustration of the proposed tree-colouring principle on
a sample tree consisting of 24 nodes. The numbers written in the
nodes are the weights assigned.\label{fig1}}
\end{figure}

In Figure \ref{fig1}, a small tree consisting of 24
nodes is depicted. The tree nodes are coloured according to the
principle described: each node is recursively assigned a
Hue interval that is proportional to its weight; Hue assigned to a
given tree node equals to the middle of this interval; Saturation
and Brightness is proportional to the distance from the root.
Each node is labelled by its weight (see Equation \eqref{eq:TreeColouring1}).
$E = 0.2$.

\section{Results}

\subsection{Example 1.: Sequence-based clustering the SwissProt database -- verification based on NCBI taxonomy identifiers}

\begin{figure}[!h]
\includegraphics[width=4.7in]{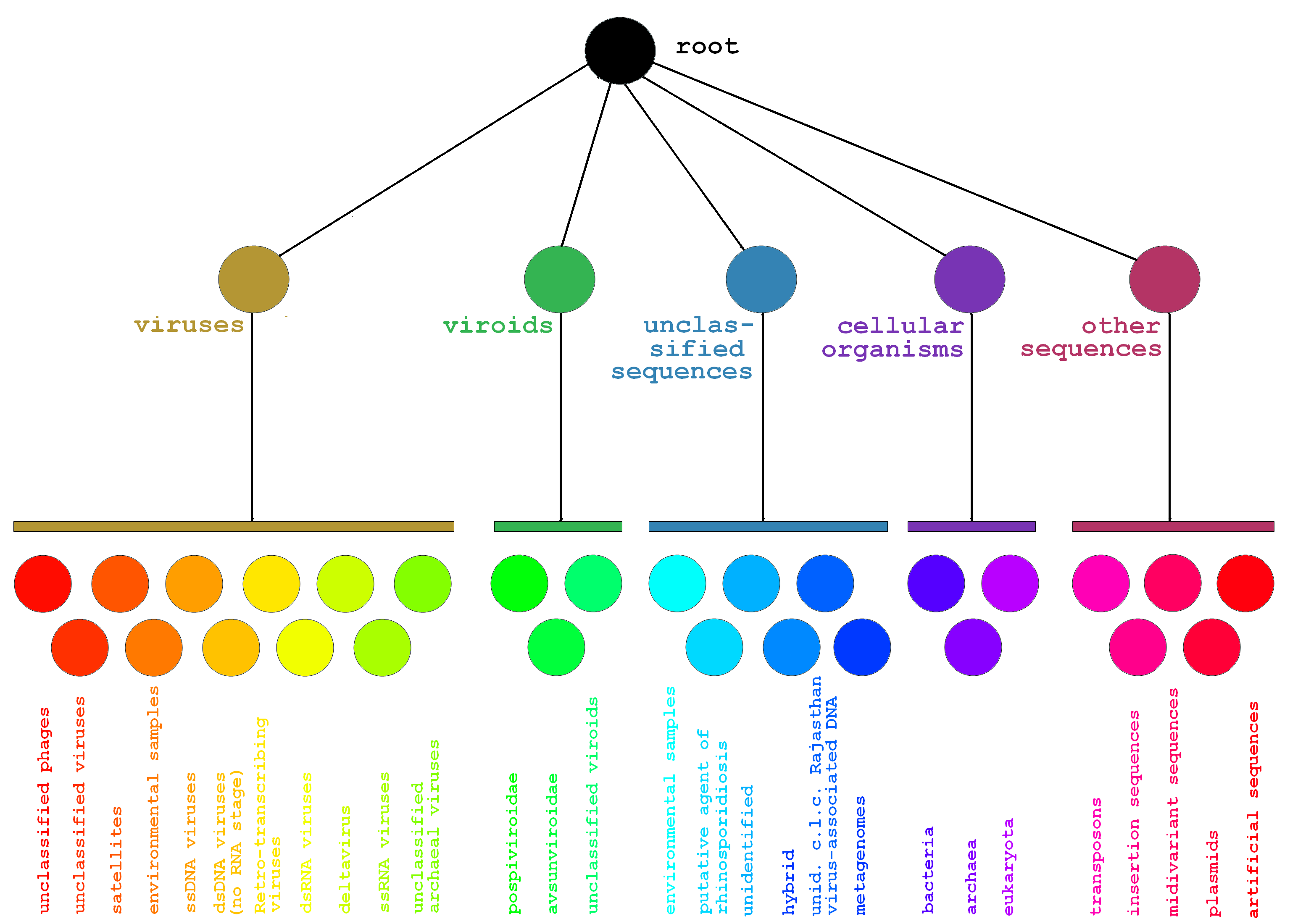}
\caption{Upper two levels of the NCBI taxonomy tree coloured by the proposed method.
Note that the colours will be different when colouring the whole tree, as nodes on
level 2 are going to have quite different weights.\label{fig0}}
\end{figure}

We applied the OPTICS algorithm to 389046 amino acid sequences occurring in
SwissProt release 55.1. In this case, the distance measure for two objects (amino acid sequences)
was based on local sequence similarity. Our goal was to visualize the species-composition
of potential clusters. To achieve this, we used the NCBI taxonomy hierarchy
(which is essentially a tree) as an \emph{a-priori} given hierarchical classification,
restricted it to the taxonomy identifiers occurring
in SwissProt\footnote{For the taxonomy-tree to remain connected, the ancestors
of these identifiers also had to be included.}, and coloured the resulting taxonomy-tree
using the method proposed. As an example, the first two levels of the NCBI taxonomy
hierarchy is shown on Figure \ref{fig0} with a possible colouring.

\begin{figure}[!h]
\includegraphics[width=4in]{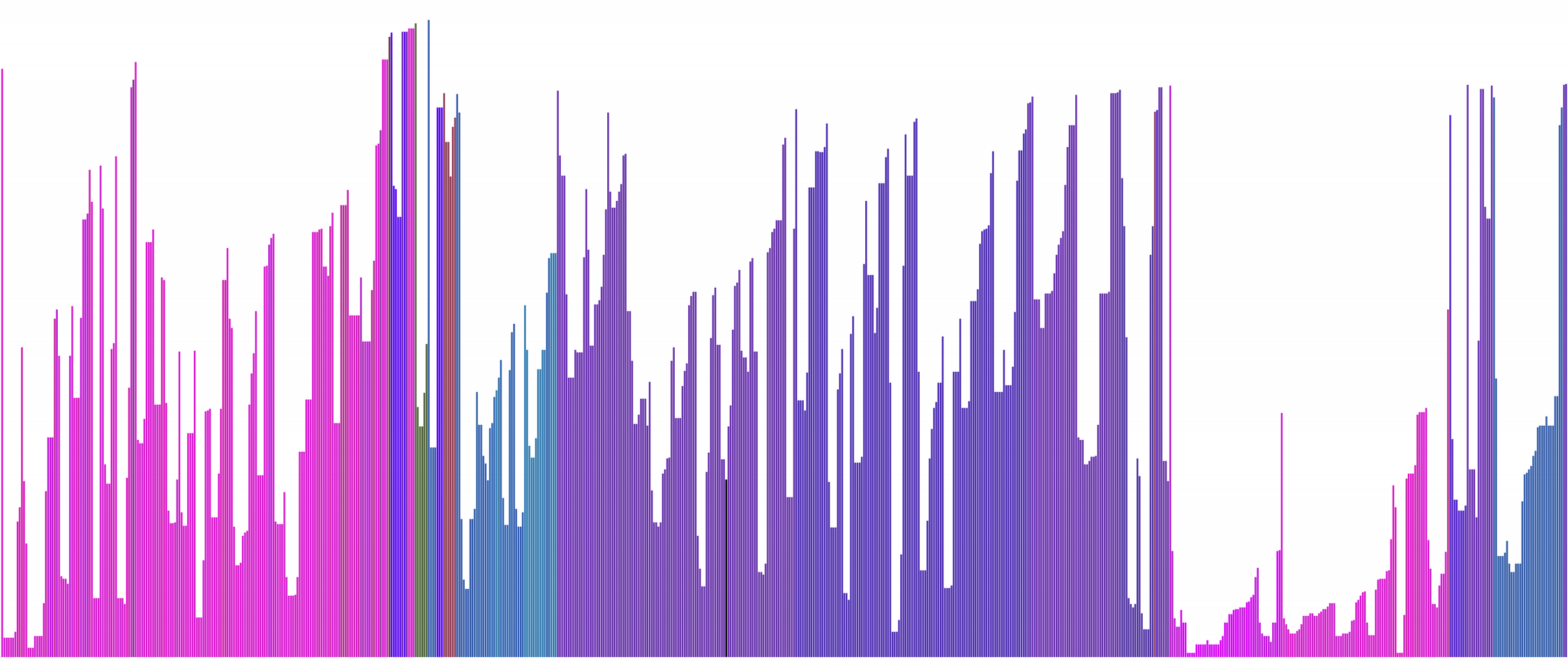}
\caption{The reachability plot of 700 sequences chosen from SwissProt, coloured by their species\label{fig4}}
\end{figure}

Due to size limitations (the x-axis of the full reachability plot consists of
the 389046 sequences being clustered), viewing the coloured reachability plot is quite a challenge.
A small, yet illustrative chunk (width: 700 sequences) of the reachability plot is shown in Figure \ref{fig4}.
On the other hand, the annotated reachability plot of the first 20000 sequences can also be
viewed on the author's home page:\\
http://www.cs.elte.hu/$ \sim $hugeaux/swissclust/swissclust\_optics\_M004\_No001.html
\footnote{This is only the first part of the reachability plot containing 20000
sequences. However, due to its substantial size, it may take some time to load properly.}

\subsection{Example 2.: Clustering locations of specific atoms in the serine protease enzyme family -- verification based on SCOP classification}

In a previous study by Ivan et al. \cite{Ivan2009}, strong correlation
has been shown to exist between the enzymatic function of serine proteases
and the orientation of amino acid side chains constituting the ''catalytic
machinery'' of the enzyme. The method described heavily depends
on the OPTICS algorithm combined with the hereby proposed visualization technique.
The main hypothesis was that some specific families
of serine proteases can be distinguished solely by taking the coordinates
of four specific atoms per protein (thus assigning a 12-dimensional
feature vector to each enzyme), and clustering these feature vectors by
the OPTICS algorithm (based on Euclidean distance as a similarity measure).
The entries (enzymes represented by their PDB \cite{Berman2003}
codes) were colored on the OPTICS reachability plot by the SCOP \cite{Murzin1995} classification
of the proteins. SCOP is a 7-level hierarchical classification of
three-dimensional protein structures; a given amino acid range of a PDB entry
belongs to exactly one SCOP class on each level of this hierarchy.
Traversing the SCOP tree from the root to a given amino acid range of a PDB entry (a leaf),
one can obtain more and more precisely defined classes the given entry belongs to.
Levels of the SCOP tree are named

\begin{itemize}
\item Root (level 0)
\item Class (level 1)
\item Fold (level 2)
\item Superfamily (level 3)
\item Family (level 4)
\item Protein (level 5)
\item Species (level 6).
\end{itemize}

The domains -- amino acid ranges -- that are classified are located on level 7.

\begin{figure}[!h]
\includegraphics[width=4in]{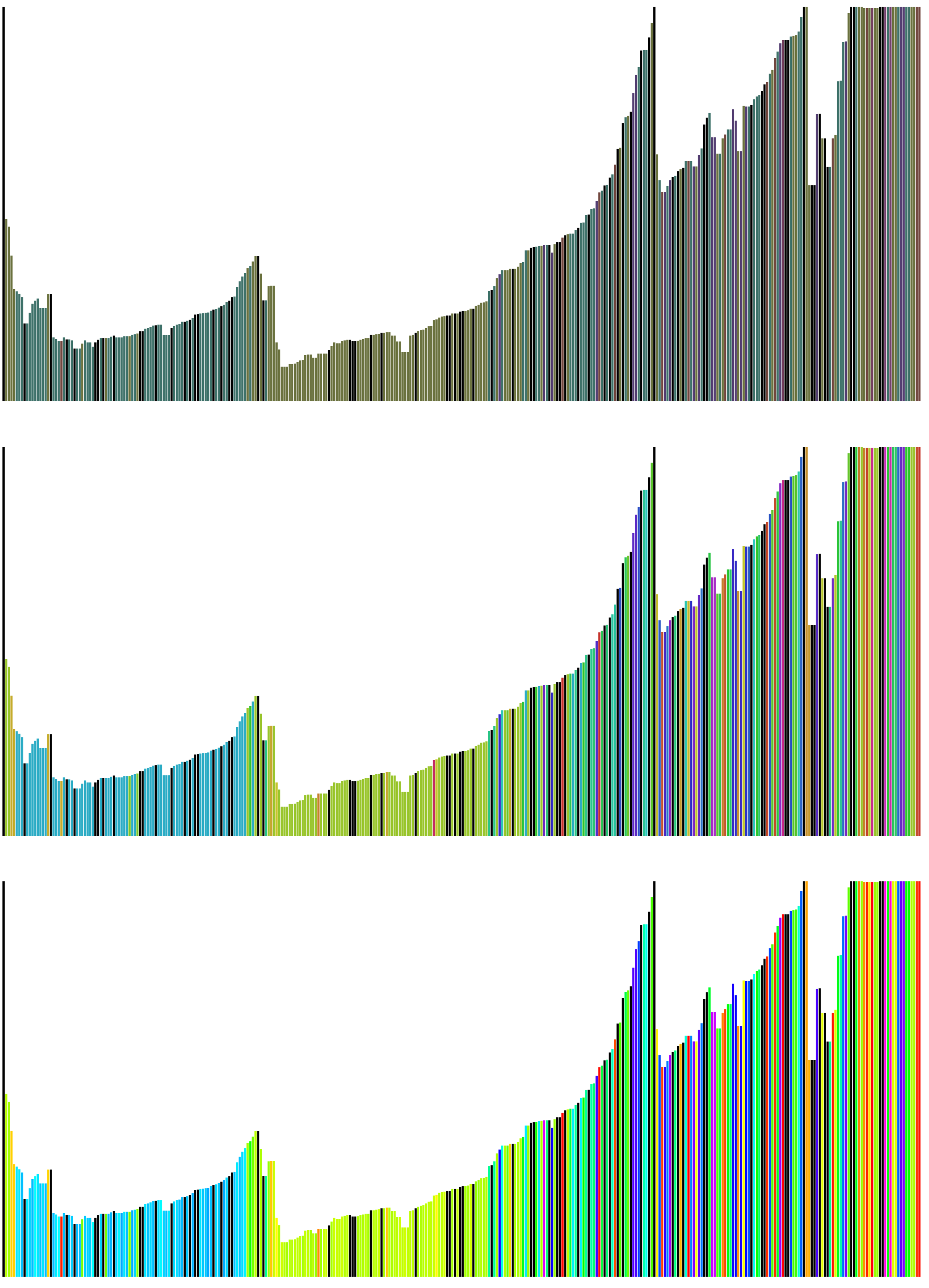}
\caption{The reachability plot of 350 PDB entries coloured by their
SCOP classification. Entries coloured black have no known SCOP classification.
Focus of this work is described in \cite{Ivan2009}.\label{fig3}}
\end{figure}

In order to visualize the assumed correlation between feature vector similarity
and SCOP classification, the 7-level SCOP hierarchy has been coloured by the
visualization method described. The two main homogenously coloured regions on the reachability
plot indicate proteins belonging to similar SCOP classes (see Figure \ref{fig3}).
The upper-panel reachability plot in Figure \ref{fig3} is coloured by the entries'
SCOP 'Class' (level 1 in SCOP tree); the reachability plot in the middle panel of
Figure \ref{fig3} is coloured by the entries' SCOP 'Superfamily' (level 3 in SCOP tree), and
the lower-panel reachability plot in Figure \ref{fig3} is coloured by the entries'
SCOP 'Protein' (level 5 in SCOP tree).

It can be seen very clearly that moving away from the root of SCOP (i.e., from the upper panel to the lower panel 
on Figure \ref{fig3}, that is, from colouring according to level 1 in the SCOP tree through the colouring by level 5 in SCOP tree at the bottom panel), colours become more saturated,
and, additionally, more and more details become visible due to subtle changes in the entries' colours, witnessing that specific 
families of serine proteases can be distinguished solely by taking the coordinates 
of four specific atoms per protein.

\section{Conclusion}

We proposed a method capable of visualizing connection between an \emph{a priori}
given classification and the OPTICS clustering algorithm's output for a given set
of input objects. We have shown how this method can be used to investigate correlations between
clustering results and \emph{a-priori} available knowledge, and also cited an example
of how the method can be used to help validating hypotheses in two bioinformatics-related tasks.
We demonstrated that the presented method is capable to visualize the correlation of the 
\emph{a-priori} available knowledge and the knowledge gained from clustering.

\end{document}